# Influences of galaxy interactions on AGN activity

Xin-Fa Deng[1]   Guisheng Yu[2]   Peng Jiang[1]

[1] School of Science, Nanchang University, Jiangxi, China, 330031

[2] Department of Natural Science, Nanchang Teachers College, Jiangxi, China, 330103

**Abstract** Using two volume-limited Main galaxy samples of the Sloan Digital Sky Survey Data Release 7 (SDSS DR7), we explore influences of galaxy interactions on AGN activity. It is found that in the faint volume-limited sample, paired galaxies have a slightly higher AGN fraction than isolated galaxies, whereas in the luminous volume-limited sample, an opposite trend can be observed. The significance is $<1\sigma$. Thus, we do not observe strong evidence that interactions or mergers likely trigger the AGN activity.

1 INTRODUCTION

Do interactions or mergers trigger activity in the nucleus of the galaxy? Many theoretical models suggest that the active galactic nucleus (AGN) activity is closely linked to galaxy interactions and mergers (e.g., Kauffmann & Haehnelt 2000; Cattaneo 2001; Wyithe & Loeb 2002; Di Matteo et al. 2003; Kang et al. 2005; Bower et al. 2006; Croton et al. 2006). But observational studies have yielded contradictory results (e.g., Dahari 1984, 1985; Keel et al. 1985; Kennicutt et al. 1987; Barton et al. 2000; Virani et al. 2000; Schmitt 2001; Miller et al. 2003; Grogin et al. 2005; Waskett et al. 2005; Koulouridis et al. 2006; Serber et al. 2006; Alonso et al. 2007; Woods & Geller 2007; Ellison et al. 2008; Li et al. 2006, 2008). Dahari (1984) searched for close companion galaxies in a redshift-limited sample of Seyfert galaxies, and found that there is a definite excess of companions in the Seyfert sample, compared with a control sample of field galaxies. Woods & Geller (2007) detected a significantly increased AGN fraction in the pair galaxies compared to matched sets of field galaxies. Ellison et al (2011) found a clear increase in the AGN fraction in close pairs of galaxies relative to the control sample, and further demonstrated that the increase in AGN fraction is strongest in equal mass galaxy pairings, and weakest in the lower mass component of an unequal mass pairing. However, Li et al. (2006) did not observe strong evidence that interactions and mergers are playing a significant role in triggering the AGN activity, and suggested other physical mechanism responsible for explaining AGN activity: AGN are preferentially located at the centres of dark matter haloes. Ellison et al. (2008) found little evidence for increased AGN activity in their close-pairs sample and concluded that, if AGN are induced by mergers, then they must occur at stages later than close-pairs typically examine. Li et al. (2008) also failed to find any corresponding relation between enhanced AGN activity and interactions.

In this study, we use the Main Galaxy sample of the Sloan Digital Sky Survey Data Release 7 (SDSS DR7) (Abazajian et al. 2009) and a relatively new and publicly available catalog of the flux and error, and explore influences of galaxy interactions on AGN activity. Close paired galaxies often be defined as interacting and merging galaxies, and are used to study the effect of galaxy interactions (e.g., Lambas et al. 2003; Alonso et al. 2004a). On the contrary, isolated galaxies are a group of galaxies which may have experienced no major interactions in billions of years. Undoubtedly, the comparison between the properties of galaxies in pairs and isolated is a useful method to unveil the effects of interactions on AGN activity.

Our paper is organized as follows. In section 2, we describe the data used. In section 3, we investigate the AGN fraction of galaxies in pairs and isolated. Our main results and conclusions are summarized in section 4.

In calculating the distance we used a cosmological model with a matter density $\Omega_0 = 0.3$, cosmological constant $\Omega_\Lambda = 0.7$, Hubble's constant $H_0 = 70 \text{km} \cdot \text{s}^{-1} \cdot \text{Mpc}^{-1}$.

**2. Data**

2.1 Summary of the data

Many of survey properties of the SDSS were discussed in detail in the Early Data Release paper (Stoughton et al. 2002). In this study, we use the Main galaxy sample (Strauss et al. 2002) of the SDSS DR7(Abazajian et al. 2009). The data were downloaded from the Catalog Archive Server of SDSS Data Release 7 by the SDSS SQL Search (with SDSS flag: bestPrimtarget&64>0) with high-confidence redshifts (Zwarning $\neq$ 16 and Zstatus $\neq$ 0, 1 and redshift confidence level: zconf>0.95) (http://www.sdss.org/dr7/).

The SDSS Main galaxy sample is an apparent-magnitude limited sample, which seriously suffers from the Malmquist bias (Malmquist 1920; Teerikorpi 1997). Because faint galaxies at large distances will not be detected, the averaged luminosity of galaxies in such a sample increases with increasing distance. In order to decrease this bias, one often used the volume-limited galaxy sample. Our Main galaxy sample of the SDSS DR7 contains 565029 Main galaxies with the redshift $0.02 \leq z \leq 0.2$. From this apparent-magnitude limited Main galaxy sample, Deng (2010) constructed a luminous volume-limited Main galaxy sample which contains 120362 galaxies at $0.05 \leq z \leq 0.102$ with $-22.5 \leq M_r \leq -20.5$ and a faint volume-limited sample which contains 33249 galaxies at $0.02 \leq z \leq 0.0436$ with $-20.5 \leq M_r \leq -18.5$. In this work, we still use these two volume-limited samples.

2.2 Galaxy pairs

For the identification of galaxy pairs, many authors developed different criteria (e.g., Karachentsev 1972; Barton et al. 2000; Lambas et al. 2003; Patton et al. 2005; Focardi et al. 2006; Kewley et al. 2006; Deng et al. 2008a, 2008b). It is important to recognize that up to now, there still is no a widely accepted criterion. Any criterion has its own drawbacks. We noted that for many issues of galaxy pairs, the use of different criteria often can reach the same conclusions. Thus, we believe that the selection of criteria are less important in such issues. In this study, we use a typical criterion developed by Lambas et al. (2003). Lambas et al. (2003) selected galaxy pairs in the field by radial velocity ($\Delta V \leq 350 \text{ km/s}$) and projected separation ($r_p \leq 100 \text{kpc}$) criteria. $r_p \leq 100$ kpc and $\Delta V \leq 350$ km/s can be defined as reliable upper limits for the relative radial velocity and projected distance criteria to select galaxy pairs with stronger specific star formation than the averaged galaxies in the SDSS and 2dF galaxy redshift survey(Lambas et al. 2003; Alonso et al. 2004b, 2006). By applying the same selection criteria, we identify 1654 pairs in the luminous volume-limited sample and 1133 pairs in the faint volume-limited sample.

It has been known for a long time that the fiber collisions are main sources of incompleteness in SDSS pair catalogs. If one attempt to study the large-scale distribution of pairs, this imcompleteness of the pair sample will be a large drawback. But in this study the influence of this imcompleteness is not crucial. In addition, correcting of some incompleteness likely results in new bias. For example, Berlind et al. (2006) corrected for fiber collisions by giving each collided galaxy the redshift of its nearest neighbor on the sky (usually the galaxy it collided with). Putting collided galaxies at the redshifts of their nearest neighbors will cause some nearby galaxies to be placed at high redshift, which artificially makes their estimated luminosities very high. Therefore, We do not make efforts to correct fiber collisions.

2.3 Control sample

A control sample is constructed by randomly selecting galaxies without close companions within $r_p$ < 100 kpc and $\Delta V$ < 350 km/s. In order to investigate the effects of galaxy interactions, one often compared galaxies in pairs with isolated galaxies. Perez et al. (2009) explored how the way of building a control sample introduce biases which could affect the interpretation of results, and claimed that a suitable control sample for isolating the effects of interactions should be built by imposing constraints on redshift, stellar mass, local environment, morphology and halo mass. But if considering the correlations among galaxy properties, one should realize that imposing too many constraints also washes out the difference of galaxy properties between the control and pair samples introduced by any physical mechanism. Because the redshift is the most fundamental quantity in selection effects, in this work, the control sample is required to have the same galaxy number and the same redshift distribution as the pair sample.

3. AGN fraction of galaxies in pairs and isolated

3.1. Identification of AGNs

By considering the classical diagnostic ratios of two pairs of relatively strong emission lines, Baldwin, Phillips &Terlevich (1981, hereafter BPT) demonstrated that it is possible to distinguish AGNs from normal star-forming galaxies. We download the flux and error in the flux of four lines from http://www.mpa-garching.mpg.de/SDSS/DR7/. Like Li et al. (2006) did, AGNs are selected from the subset of galaxies with signal-to-noise ratio S/N > 3 on the four emission lines [OIII]$\lambda$5007, H$\beta$, [NII]$\lambda$6584, H$\alpha$. In our apparent-magnitude limited Main galaxy sample, 253594 galaxies have S/N >3 on the 4 lines. Following Kauffmann et al. (2003), a galaxy is defined to be an AGN if log ([OIII]$\lambda$5007/H$\beta$ )>0.61/{log([NII]$\lambda$6584/H$\alpha$ )-0.05}+1.3.

Our apparent-magnitude limited Main galaxy sample contains 89716 AGNs. Fig.1 shows redshift distribution of Main galaxies and AGNs for the apparent-magnitude limited Main galaxy sample of the SDSS DR7. As seen form this figure, there is a quite large difference of AGN fraction in different redshift bins, but the trend of change for AGNs is nearly the same as the one for Main galaxies. This shows that there is serious selection effects in the apparent-magnitude limited Main galaxy sample. In this work, we use the volume-limited galaxy samples in which the radial selection function is approximately uniform. In addition, when performing comparative studies, the control sample is required to have the same galaxy number and the same redshift distribution as the pair sample. So, selection effects in this work are less important.

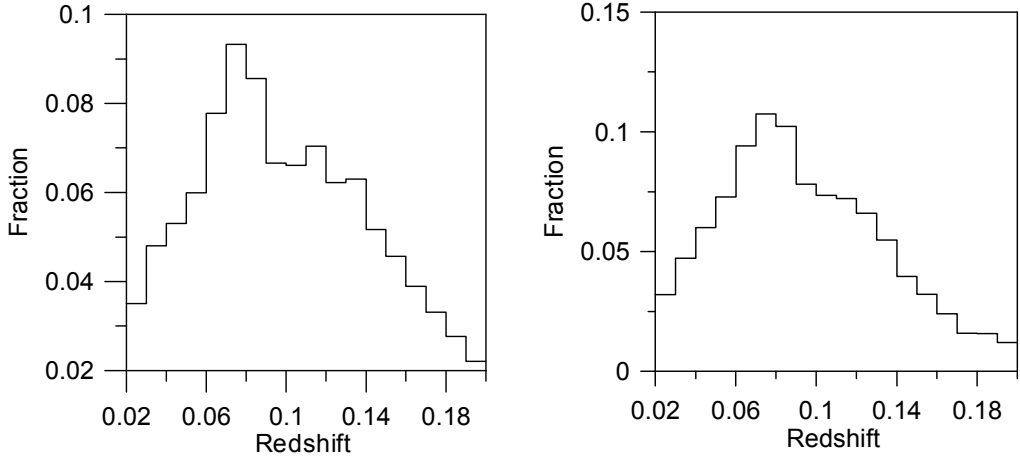

Fig.1 Redshift distribution of Main galaxies (left panel) and AGNs (right panel) for the apparent-magnitude limited Main galaxy sample of the SDSS DR7.

**3.2. AGN fraction of galaxies in pairs and isolated**

The luminous volume-limited sample contains 28674 AGNs, the faint volume-limited sample includes 5021 AGNs. We compute the AGN fraction of galaxies in pairs and isolated: in the luminous volume-limited sample, $0.2597\pm0.0089$ for isolated galaxies and $0.2358\pm0.0084$ for paired galaxies; in the faint volume-limited sample, $0.1545\pm0.0083$ for isolated galaxies and $0.1664\pm0.0086$ for paired galaxies. Here, the Poissonian error is taken into account. In the faint volume-limited sample, paired galaxies have a slightly higher AGN fraction than isolated galaxies, which seemingly shows that interactions or mergers likely trigger the AGN activity. But in the luminous volume-limited sample, an opposite trend can be observed. The significance is $<1\sigma$. So, it is difficult to conclude whether interactions or mergers likely trigger the AGN activity.

Considering the variation in AGN fraction with redshift, we divide the whole redshift region of two volume-limited samples into redshift bins with width 0.01 (The last redshift bin is 0.100-0.102 for the luminous volume-limited Main galaxy sample, and 0.04-0.0436 for the faint volume-limited Main galaxy sample), and focus the analysis on the statistical differences of the AGN fraction between paired galaxies and isolated ones in each redshift bin. Fig.2 shows the fraction of AGNs as a function of redshift z for paired galaxies (red triangle) and isolated galaxies (blue dot) in the luminous (on the right-hand side of the green vertical line) and faint (on the left-hand side of the green vertical line) volume-limited samples. As can be seen from Fig.2, in the faint volume-limited sample (low redshift range), paired galaxies have a slightly higher AGN fraction than isolated galaxies, whereas in the luminous volume-limited sample (high redshift range), the AGN fraction of isolated galaxies is slightly higher. This finding further confirms the above-mentioned conclusion.

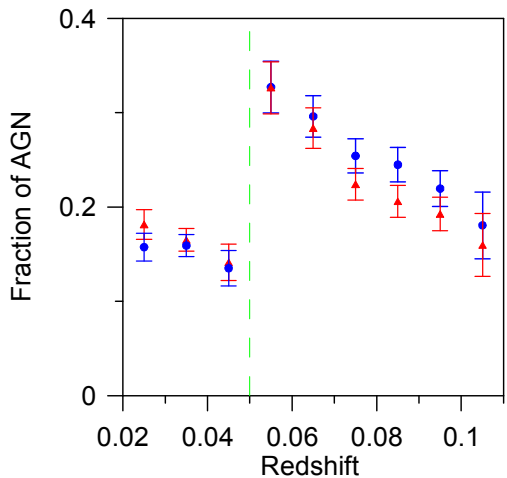

Fig.2 Fraction of AGNs as a function of redshift z for paired galaxies (red triangle) and isolated galaxies (blue dot) in the luminous (on the right-hand side of the green vertical line) and faint (on the left-hand side of the green vertical line) volume-limited samples. The error bars are 1 $\sigma$ Poissonian errors.

In dense systems of a galaxy sample, interactions and mergers often occur in a large fraction of galaxies (e.g., Rubin et al 1991; Mendes de Oliveira & Hickson 1994; Lee et al. 2004). For example, Lee et al. (2004) showed that there is strong evidence of interactions and mergers within a significant fraction of SDSS CGs (compact groups of galaxies). Paired galaxies also often be located in dense systems such as groups and clusters. Many authors showed that there is no evidence for the environmental dependence of the AGN fraction (e.g., Monaco et al. 1994; Coziol et al. 1998; Shimada et al. 2000; Carter et al. 2001; Schmitt 2001; Miller et al. 2003). For example, Carter et al. (2001) showed that the AGN fraction is insensitive to the local environment. Miller et al. (2003) also observed that this fraction is constant from the cores of galaxy clusters to the rarefied field population. There also have been a number of dissenting papers. For example, Dressler et al. (1985) found 5 times as many AGNs in the field as in clusters. Popesso & Biviano (2006) also reported a lower fraction of (weak and strong) optical AGN in clusters than in the field and smaller systems. According to these two standpoints, it is difficult to reach the conclusion: paired galaxies have a higher AGN fraction than isolated galaxies.

## 4. Summary

Using two volume-limited Main galaxy samples of the SDSS DR7, we explore influences of galaxy interactions on AGN activity. In each sample, we construct a paired sample and a control sample, and compared the AGN fraction of galaxies in pairs and isolated. The control sample is required to have the same galaxy number and the same redshift distribution as the pair sample. It is found that in the faint volume-limited sample, paired galaxies have a slightly higher AGN fraction than isolated galaxies, whereas in the luminous volume-limited sample, an opposite trend can be observed. The significance is $<1\sigma$. Thus, we do not observe strong evidence that interactions or mergers likely trigger the AGN activity.

## Acknowledgements

We thank the anonymous referee for many useful comments and suggestions. Our study was supported by the National Natural Science Foundation of China (NSFC, Grant 10863002).

Funding for the SDSS and SDSS-II has been provided by the Alfred P. Sloan Foundation, the Participating Institutions, the National Science Foundation, the US Department of Energy, the National Aeronautics and Space Administration, the Japanese Monbukagakusho, the Max Planck Society, and the Higher Education Funding Council for England. The SDSSWeb site is http://www.sdss.org.

The SDSS is managed by the Astrophysical Research Consortium for the Participating Institutions. The Participating Institutions are the American Museumof Natural History, Astrophysical Institute Potsdam, University of Basel, University of Cambridge, Case Western Reserve University, University of Chicago, Drexel University, Fermilab, the Institute for Advanced Study, the Japan Participation Group, Johns Hopkins University, the Joint Institute for Nuclear Astrophysics, the Kavli Institute for Particle Astrophysics and Cosmology, the Korean Scientist Group, the Chinese Academy of Sciences (LAMOST), Los Alamos National Laboratory, the Max Planck Institute for Astronomy (MPIA), the Max Planck Institute for Astrophysics (MPA), New Mexico State University, Ohio State University, University of Pittsburgh, University of Portsmouth, Princeton University, the US Naval Observatory, and the University of Washington.